\documentclass[letterpaper,journal]{IEEEtran}
\usepackage{amsmath,amsfonts,amssymb}
\usepackage{bm}
\usepackage{algorithm}
\usepackage{algorithmic}
\usepackage{array}
\usepackage{textcomp}
\usepackage{stfloats}
\usepackage{url}
\usepackage{verbatim}
\usepackage{graphicx}
\usepackage{cite}
\usepackage{multicol}
\usepackage{multirow}
\usepackage{breqn}
\usepackage{booktabs}
\usepackage{makecell}
\usepackage{color}
\usepackage{soul}
\usepackage[font=normalsize,labelfont=bf,labelsep=colon,justification=centering]{caption}
\usepackage[font=normalsize,labelfont=normalfont,textfont=normalfont,justification=centering]{subfig}
\begin{document}

\title{\fontsize{23pt}{27pt}\selectfont Constrained Pinching Antenna Array Design for Sum-Rate Maximization in Multi-User PASS}

\author{Minghao Jin, Anna Li,~\IEEEmembership{Member,~IEEE}, Tianwei Hou,~\IEEEmembership{Member,~IEEE}, \\ Qiang Ni,~\IEEEmembership{Senior Member,~IEEE}, Arumugam Nallanathan,~\IEEEmembership{Fellow,~IEEE}
\thanks{Minghao Jin, Anna Li and Qiang Ni are with the School of Computing and Communications, Lancaster University, Lancaster LA1 4WA, U.K. (e-mail: m.jin6@lancaster.ac.uk; a.li16@lancaster.ac.uk; q.ni@lancaster.ac.uk).}
\thanks{Tianwei Hou is with the School of Electronic and Information Engineering, Beijing Jiaotong University, Beijing 100044, China (e-mail: twhou@bjtu.edu.cn).}
\thanks{Arumugam Nallanathan is with the School of Electronic Engineering and Computer Science, Queen Mary University of London, E1 4NS London, U.K., and also with the Department of Electronic Engineering, Kyung Hee University, Yongin-si, Gyeonggi-do 17104, South Korea (e-mail: a.nallanathan@qmul.ac.uk).}
}



\maketitle

\begin{abstract}
Pinching antenna systems (PASS) have recently emerged as a promising architecture for flexible indoor wireless communications. However, most existing pinching antenna (PA) array designs for multi-user PASS either offer limited beam adaptation accuracy or require prohibitively high deployment cost. In this paper, we investigate a more practical constrained pinching antenna array (C-PAA)-assisted downlink PASS, where multiple PAs are grouped into a movable array and can be finely adjusted within the array at the wavelength scale.
To improve the system spectral efficiency, a sum-rate maximization problem is formulated by jointly considering the array-center position and the fine-grained antenna distribution within the C-PAA. First, the structural properties of the C-PAA are characterized, and an explicit upper bound on the array aperture is derived. Then, tractable approximations for the effective channel gain and the achievable user rate are developed.
Furthermore, the optimization problem of the multi-user sum-rate is analyzed, where the system sum-rate function is shown to exhibit a favorable unimodal behavior under practically relevant conditions, which enables an efficient one-dimensional search for the optimal C-PAA position. To further reduce the computational complexity, a closed-form approximate solution for the near-optimal array-center position is derived. Numerical results verify the accuracy of the developed analysis and demonstrate that the proposed C-PAA scheme closely approaches the ideal upper bound and significantly outperforms conventional fixed-spacing and existing PA array benchmarks.
\end{abstract}

\begin{IEEEkeywords}
array gain, PASS, pinching antenna array, performance analysis, sum rate optimization.
\end{IEEEkeywords}

\section{Introduction}

\IEEEPARstart{W}{ith} the rapid evolution of wireless networks from 5G toward 6G, future communication systems are expected to support not only higher spectral efficiency but also greater spatial flexibility and adaptability. Conventional fixed-antenna architectures, although highly successful in previous generations, are becoming increasingly inadequate for emerging indoor scenarios that require agile beam control, low deployment cost, and high energy efficiency \cite{Larsson2014MassiveMIMO, Boccardi2014FiveDirections, Saad2020Vision6G}. To overcome the limitations of rigid antenna structures, a variety of advanced electromagnetic and antenna technologies have been developed, including reconfigurable intelligent surfaces (RIS), holographic or continuous-aperture architectures, and other programmable radio environments \cite{Liu2021RIS, Gong2024NearFieldHolo, Sanguinetti2023WDMHolo, HouCAPAOptimization}. These technologies can significantly enhance wireless capability by reshaping the propagation environment or enlarging the effective aperture of the transceiver.

Beyond electromagnetic programmability, another important research direction is to improve communication performance through the spatial flexibility of the antenna itself. Representative examples include movable antennas and fluid antenna systems, which enable adaptive repositioning of radiating elements to exploit favorable propagation conditions \cite{Zhu2025MATutorial, New2025FASTutorial, Wong2020FluidAntenna}. Nevertheless, the performance gains of these techniques are often closely tied to small-scale fading variations, and their advantages may diminish in line-of-sight (LoS)-dominant indoor environments. This motivates the study of pinching antenna systems (PASS), where dielectric pinches are flexibly deployed along a waveguide to radiate or receive electromagnetic signals at desired locations \cite{Liu2025PASSPerspective, Fukuda2022PinchingAntenna}. Owing to the unique structure, PASS can directly manipulate effective radiation positions along the waveguide and therefore provide a new mechanism for mitigating large-scale path loss in indoor wireless communications \cite{Ding2025FlexibleAntenna}.

Existing studies have demonstrated the considerable potential of PASS for indoor wireless communications. In particular, the fundamental architecture and opportunities of PASS were introduced in \cite{Ding2025FlexibleAntenna}, while the array-gain of multiple pinching antennas deployed on a single waveguide was characterized in \cite{Ouyang2025ArrayGain}. Building on these foundations, several advanced PASS transmission designs have been developed from different perspectives, including antenna activation and multiple-access design for PASS \cite{Wang2025ActivationNOMA, Ding2025EDMA, Tegos2025UplinkRSMA, Xie2025LowComplexityPlacement}, downlink beamforming with PA assisted MIMO systems \cite{Bereyhi2025DownlinkBeamforming, P6}, and PASS-enabled integrated sensing and communication (ISAC) \cite{I1, I2, I3, I4}.
For multi-user and multiple-access scenarios, prior studies have investigated PASS from several complementary perspectives, including throughput and fairness optimization \cite{Che2026ThroughputTDMA, Tegos2025MinRateUplink, Oikonomou2025OFDMA4PASS}, capacity characterization \cite{Xiao2025PASSMultiUserCapacity, Chen2026SumCapacityMAC, Ouyang2025CapacityPASS, Hou2026ULPASS}, and richer transmission architectures, such as waveguide division multiple access, OMA/NOMA comparison, and multi-waveguide transmission with joint beamforming and resource allocation \cite{Zhao2025WDMA, Ren2025PASSNOMAOMA, Zhao2025MultiWaveguidePASS, Wang2026AnActivationMWPASS}. Collectively, these studies demonstrate that PASS can effectively exploit spatial reconfigurability to improve large-scale channel conditions and thereby substantially enhance multi-user communication performance.

From the perspective of practical implementation, however, existing multi-user PA deployment strategies can be broadly classified into two categories according to whether the PA positions are fixed for all users or reconfigured for the currently served user. 
In the first category, the PA locations are optimized only once based on the locations of all users in the considered multi-user scenario, and the resulting PA configuration is then fixed for serving all users without real-time adjustment to the currently scheduled user \cite{Xiao2025PASSMultiUserCapacity, Tegos2025MinRateUplink, Oikonomou2025OFDMA4PASS, Ouyang2025CapacityPASS}.
Although such a design is easy to deploy in practice and can balance the performance of multiple users, it does not enable precise user-specific beam pointing, thereby limiting the performance potential of PASS.

In the second category, the PA positions are adaptively configured for the currently served user. Such user-specific designs can be further divided into two main types. One is piano-based PASS, where a set of candidate PA locations is pre-installed and only selected elements are activated according to the scheduled user \cite{Xiao2025PASSMultiUserCapacity, Wang2025ActivationNOMA, Wang2026AnActivationMWPASS}. The other allows individual PAs, or PA arrays, to be repositioned toward user-dependent locations along the waveguide \cite{Ding2025FlexibleAntenna, Xie2025LowComplexityPlacement, Che2026ThroughputTDMA, Chen2026SumCapacityMAC, Hou2026ULPASS}. Although these user-specific designs can improve the received signal strength of the served user, they still face practical implementation limitations.
In particular, piano-based PASS is constrained by discrete candidate locations and therefore offers limited beam adaptation resolution. Unconstrained user-specific PA repositioning is also difficult to implement in practice: it either requires a dedicated PA or PA set for each user, which is costly and may cause physical overlap in dense scenarios, or relies on instantaneous PA movement across users, which is unrealistic due to mechanical reconfiguration delay and complexity. Therefore, despite its theoretical advantages, unconstrained PA repositioning is difficult to scale in dense and dynamic indoor deployments.
To address these issues, a pinching antenna array architecture, in which multiple PAs are grouped into an independent movable module, was proposed in \cite{houPerformancePAA}, where analytical expressions for the antenna gain and achievable rate were also derived.
However, due to the fixed-spacing constraint imposed on the PA elements, that work mainly focuses on performance characterization and priority-oriented transmission, and does not provide a comprehensive performance optimization framework for multi-user scenarios.

\subsection{Motivation and Contribution}

From the above discussion, it is clear that most existing studies have primarily focused on theoretical optimal PA deployment, while relatively limited attention has been devoted to cost-efficient and practically constrained array architectures. Although the constrained pinching antenna array concept has been introduced in \cite{houPerformancePAA}, its system-level performance optimization remains largely unexplored. This gap motivates the following key research questions: \textit{i)} how to design a constrained pinching antenna array that can flexibly support users with heterogeneous and dynamically varying locations, while properly balancing the physical aperture constraint and the beamforming capability; \textit{ii)} how to accurately characterize and efficiently optimize the performance of such a constrained PASS architecture, ideally with low-complexity algorithms or even closed-form solutions; and \textit{iii)} how much performance is lost due to the array-size constraint, and whether the resulting design remains robust under different deployment scenarios.

Motivated by these observations, this paper investigates the performance optimization of constrained pinching antenna systems. Different from existing PASS models, the considered framework explicitly accounts for the physical aperture of the constrained pinching antenna array (C-PAA) and studies the joint effect of array-center movement and intra-array wavelength-scale adjustment. The main contributions of this paper are summarized as follows:
\begin{itemize}
    \item We propose a C-PAA architecture for downlink PASS, in which multiple PAs are grouped into a movable array deployed on a ceiling-mounted waveguide. Based on this architecture, a practical sum-rate maximization problem is formulated by jointly optimizing the array-center position and the fine-grained PA positions within the C-PAA.
    
    \item We characterize the C-PAA structure under the adjacent-PA phase-alignment condition and derive an explicit upper bound on the array aperture. Based on the proposed phase-alignment design, tractable approximations for the effective channel gain and the achievable rate are further developed.
    
    \item We analyze the multi-user sum-rate optimization problem and demonstrate that, although the individual user rate is not globally concave, the expected sum-rate function remains strictly concave under uniform user distributions. Based on this, we develop an efficient one-dimensional search algorithm and further derive a closed-form approximate solution with reduced computational complexity.
    
    \item Numerical results verify the accuracy of the proposed analytical approximations and demonstrate that the C-PAA can closely approach the ideal upper bound while consistently outperforming conventional fixed-spacing and priority-oriented benchmark schemes.
\end{itemize}

\subsection{Organization}

The remainder of this paper is organized as follows. Section II introduces the system model and formulates the sum-rate maximization problem. Section III presents the C-PAA design together with the corresponding channel statistics and achievable-rate analysis. Section IV investigates the sum-rate optimization problem and develops both iterative and closed-form approximate solutions. Section V provides numerical results and performance comparisons. Finally, Section VI concludes this paper.

\section{System Model}

\subsection{Antenna and Channel Model}

\begin{figure}[t]
\centering
\includegraphics[width=1\linewidth]{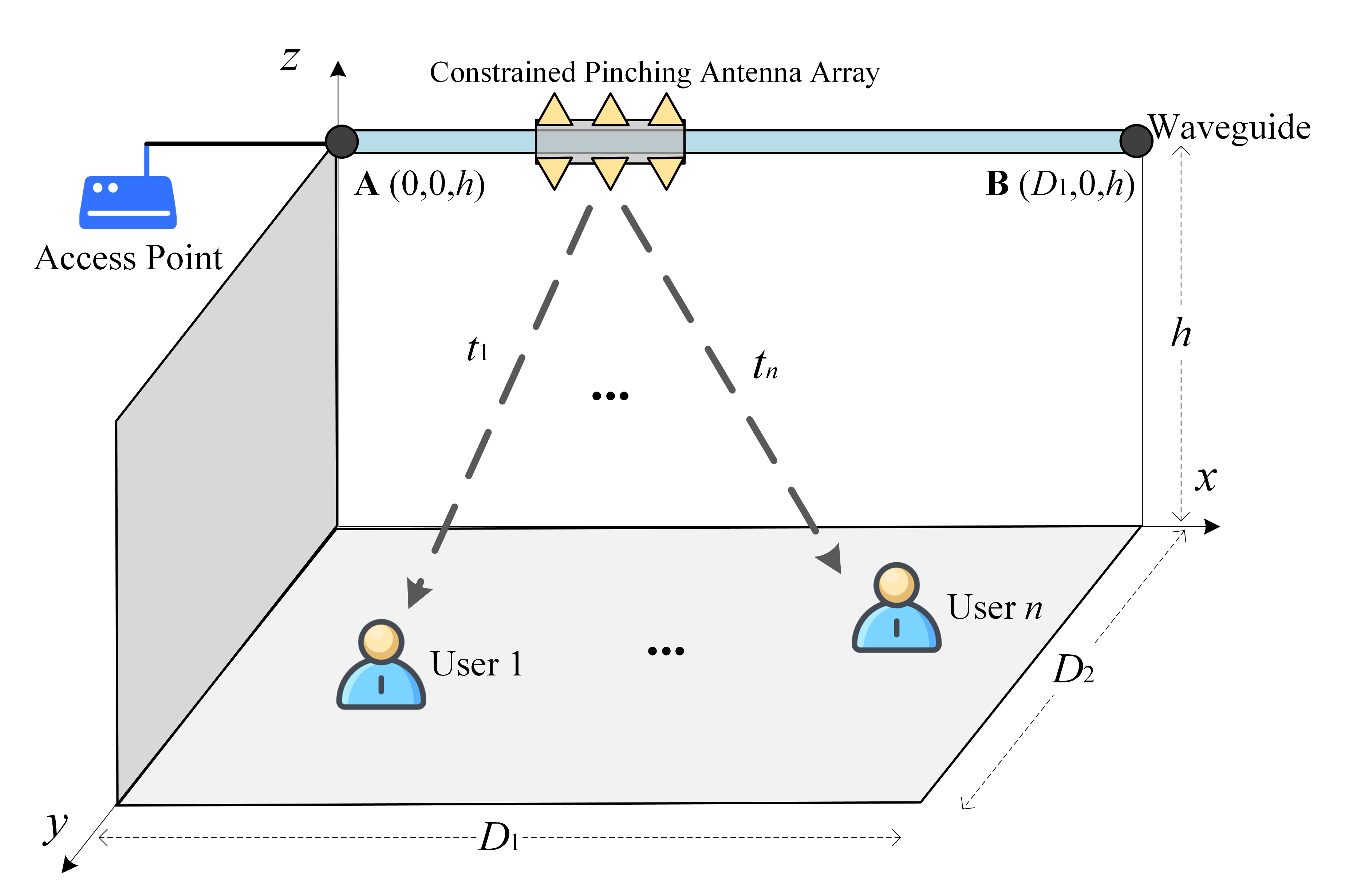}
\caption{Illustration of the considered C-PAA assisted downlink pinching antenna systems.}
\label{fig:Illustration_CPAA}
\end{figure}

As shown in Fig. \ref{fig:Illustration_CPAA}, we consider a downlink multi-user communication scenario, where a dielectric waveguide serves $N$ single-antenna users. In this paper, we assume that multiple users are randomly distributed across the floor, whose locations are indicated by $\mathbf{U}_n = [x_n,y_n,0]$, $1 \leq n \leq N$. Without loss of generality, time-division multiple access (TDMA) is adopted as a representative orthogonal multiple access (OMA) scheme, such that user $\mathbf{U}_n$ is scheduled in time slot $t_n$. The room dimensions are defined as $x\in[0,D_1]$, $y\in[0,D_2]$, and $z\in[0,h]$. An access point (AP) is deployed at a corner of the ceiling, and a waveguide is mounted along the ceiling edge. The start and end points of the waveguide are denoted by $\mathbf{A} = [0,0,h]$ and $\mathbf{B} = [D_1,0,h]$, respectively. Hence, an arbitrary point on the waveguide can be expressed as
\begin{equation}
\label{eq:WG}
    \mathcal{L}(r) = \mathbf{A}+(\mathbf{B}-\mathbf{A})r, \quad r\in[0,1].
\end{equation}

A total of $M$ PAs are deployed on a constrained pinching antenna array (C-PAA) that can move along the waveguide. Moreover, each PA can be adjusted at the wavelength scale within the array to enable phase alignment of the signals from different PAs at the user side. The position of the $m$-th PA is given by $\mathbf{Q}_m(t_n) = [x_m(t_n), 0, h]$, where $m \in [1, M]$ and the position $x_m(t_n)$ is finely adjusted across different user time slots $t_n$. For simplicity, when $x_m(t_n)$ holds for all $n$, it can be denoted simply as $x_m$. Without loss of generality, the antennas are indexed in ascending order according to their $x$-coordinates, i.e., $x_1 \leq x_2 \leq \cdots \leq x_M$.

To facilitate the theoretical characterization of fundamental performance limits, we restrict our analysis to LoS propagation in the large-scale channel model.
Under the spherical-wave assumption, the LoS channel coefficient between the \textit{m}-th PA and the \textit{n}-th user is given by \cite{Ding2025FlexibleAntenna, Ouyang2025ArrayGain}
\begin{equation}
    h_{m,n} = \frac{\gamma^{\frac{1}{2}}e^{-j\frac{2\pi}{\lambda}||\mathbf{U}_n-\mathbf{Q}_m(t_n)||}}{||\mathbf{U}_n-\mathbf{Q}_m(t_n)||},
\end{equation}
where $\gamma = (\frac{c}{4\pi f_c})^2$, $c$ denotes the speed of light, $f_c$ is the carrier frequency, $\lambda$ is the free-space wavelength, and $||\cdot||$ represents the Euclidean norm, which can be expressed as:
\begin{equation}
    ||\mathbf{U}_n-\mathbf{Q}_m(t_n)|| = \sqrt{(x_n-x_m(t_n))^2 + y_n^2 + h^2}.
\end{equation}
For notational simplicity, the distance between $n$-th user and $m$-th PA is hereafter denoted as $d_{n,m} = ||\mathbf{U}_n-\mathbf{Q}_m(t_n)||$.
Neglecting the propagation loss inside the waveguide \cite{Hou2026ULPASS, Wang2025ActivationNOMA}, the phase shift introduced by the propagation distance from the AP to the $m$-th PA at slot $t_n$ is expressed as:
\begin{equation}
    \phi_{m,n} = e^{-j\frac{2\pi}{\lambda_g}||\mathbf{Q}_m(t_n)-\mathbf{A}||},
\end{equation}
where wavelength in waveguide $\lambda_g = \lambda/n_{\text{eff}}$, and $n_{\text{eff}}$ denotes the effective refractive index of the dielectric waveguide.

\subsection{Signal Model}
Let $s_n \in \mathbb{C}$ denote the normalized transmit symbol for user $n$, and let $P$ represent the total transmit power. Owing to the characteristics of the waveguide and PAs, equal power allocation among all PAs is assumed. Accordingly, each antenna is allocated a transmit power of $\frac{P}{M}$, and the received signal at the $n$-th user can be expressed as
\begin{equation}
    y_n = \sum_{m=1}^{M}\sqrt{\frac{P}{M}}h_{m,n}\phi_{m,n} s_n + \eta_n,
\end{equation}
where $\eta_n \sim \mathcal{N}(0,\sigma^2)$ denotes the additive white Gaussian noise (AWGN) with variance $\sigma^2$. Therefore, the achievable data rate of user $n$ is given by
\begin{equation}
    R_n=\frac{1}{N}\log_2\Big(1+\frac{P}{M\sigma^2}\Big|\sum_{m=1}^M h_{m,n}\phi_{m,n} \Big|^2 \Big).
\end{equation}

\subsection{Sum Rate Problem Formulation}

The achievable data rates of users are highly dependent on the positions of the PAs. In particular, there exists an optimal deployment of the PAs on the C-PAA, denoted by $\{\mathbf{Q}^*_m\}_{m=1}^M$.
To improve the system performance, we formulate the sum-rate maximization problem with respect to the PA positions as follows:
\begin{subequations}\label{probSumRate}
\begin{align}
\max_{\{\mathbf{Q}_m\}_{m=1}^M} \quad
& R_{\mathrm{sum}} = \sum_{n=1}^{N} R_n, \label{7a}\\
\text{s.t.} \quad
& \mathbf{Q}_{\text{C-PAA}} \in \mathcal{L}, \label{7b}\\
& \mathbf{Q}_m \in \mathcal{Q}(\mathbf{Q}_{\text{C-PAA}}), \quad \forall m, \label{7c}
\end{align}
\end{subequations}
where constraint \eqref{7b} ensures that the C-PAA is deployed along the waveguide $\mathcal{L}$, while \eqref{7c} enforces that each PA lies within the spatial extent of the C-PAA, i.e., $\mathcal{Q}(\mathbf{Q}_{\text{C-PAA}})$ denotes the feasible region determined by the C-PAA structure.

\section{Performance Analysis}
\subsection{Constrained Pinching Antenna Array Design}
In this subsection, we investigate the design of the constrained pinching antenna array (C-PAA). To characterize its position along the waveguide, we introduce a reference center $\mathbf{Q}_c=[x_c,0,h]$. Specifically, symmetrically located PAs on the two sides of $\mathbf{Q}_c$ have identical wavelength differences. When $M$ is odd, $\mathbf{Q}_c$ coincides with the central PA location, i.e., $\mathbf{Q}_c=\mathbf{Q}_{(M+1)/2}$. Note that $\mathbf{Q}_c$ serves as an optimization-oriented reference point, which does not necessarily coincide with the geometric center of the resulting PA positions for a given served user. Accordingly, the $x$-coordinate of the $m$-th PA when serving user $n$ can be expressed as

\begin{equation}
\label{x_m}
    x_m(t_n)=x_c+v_m(t_n),
\end{equation}
where $v_m(t_n)$ denotes the offset of the $m$-th PA relative to $x_c$ during the time slot serving user $n$. Similarly, when $v_m(t_n)$ holds for all $n$, it can be simplified as $v_m$.

To minimize the physical size of the C-PAA and reduce the movement distance of PAs during user switching, we assume that the phase difference between the signals transmitted by adjacent PAs toward the user is $2\pi$. Under this condition, $v_m$ should satisfy:
\begin{equation}
\label{v_m}
    \frac{v_m(t_n)}{\lambda_g}+\frac{d_{n,m}}{\lambda}=\frac{d_{n,c}}{\lambda}+m-\frac{M+1}{2},
\end{equation}
where $v_m(t_n)$ can be obtained by solving (\ref{v_m}) with a given value of $x_c$ and $d_{n,c} = ||\mathbf{U}_n-\mathbf{Q}_c||$.

\noindent\textbf{Proposition 1.} 
Based on the phase-alignment PA positions given in (\ref{v_m}), the aperture size of the C-PAA satisfies:
\begin{equation}
\label{aperture}
    x_M-x_1<\frac{(M-1)\lambda\lambda_g}{\lambda-\lambda_g}.
\end{equation}

\noindent\textit{Proof.} 
From (\ref{x_m}), for the time slot serving any user $n$, the array aperture can be expressed as $x_M - x_1 = v_M - v_1$. By substituting $m = 1$ and $m = M$ into (\ref{v_m}), taking the difference, and performing straightforward algebraic manipulations, we obtain:
\begin{equation}
\label{vM-v1}
    v_M-v_1=(M-1)\lambda_g+\frac{\lambda_g}{\lambda}(d_{n,1}-d_{n,M}).
\end{equation}
By applying the triangle inequality, it follows that 
\begin{equation}
\label{ineq:vM-v1}
    |d_{n,1} - d_{n,M}| < v_M - v_1, \quad \forall n.
\end{equation}
When $n_{\text{eff}}>1$, it follows that $\lambda>\lambda_g$. By substituting (\ref{ineq:vM-v1}) into (\ref{vM-v1}) and applying simple algebraic transformations, (\ref{aperture}) is obtained, which completes the proof.

\noindent\textbf{Remark 1.}
Under the adjacent-PA phase-alignment condition, the aperture of the C-PAA admits a user-independent upper bound. As a result, the array remains physically compact even when the PA positions are adaptively reconfigured across user time slots, which also facilitates practical user switching within the C-PAA.

\subsection{Channel Statistics and Achievable Rate}

With the PA positions finely adjusted on the C-PAA, the signals radiated by different PAs can be coherently aligned at the user side. Accordingly, the effective array gain is determined by the distances between the PAs and the user. When $M$ PAs are optimally clustered along the waveguide, the effective channel gain for ground user $n$ can be expressed as
\begin{equation}
\label{ArrayGainSum}
\begin{split}
    \varepsilon_n 
    &= \sum_{m=1}^M \frac{1}{\|\mathbf{U}_n-\mathbf{Q}_m(t_n)\|} \\
    &= \sum_{m=1}^M \frac{1}{\sqrt{(x_m(t_n)-x_n)^2+y_n^2+h^2}} \\
    &= \sum_{m=1}^M 
    \frac{1}{\sqrt{\big(x_c-x_n+v_m(t_n)\big)^2+y_n^2+h^2}}.
\end{split}
\end{equation}

Since the discrete summation in \eqref{ArrayGainSum} does not lead to a compact and analytically convenient expression for subsequent derivations, we resort to an approximation that enables more tractable analysis.

\noindent\textbf{Lemma 1.}
Under the phase-alignment distribution of $M$ PAs, the effective channel gain $\varepsilon_n$ can be approximated as follows:
\begin{equation}
\label{ArrayGainResult}
    \varepsilon_n 
    \approx \frac{1}{\lambda_g}
    \ln \left(
    \frac{\sqrt{(x_c-x_n+a)^2+y_n^2+h^2}+x_c-x_n+a}
    {\sqrt{(x_c-x_n-a)^2+y_n^2+h^2}+x_c-x_n-a}
    \right),
\end{equation}
where $a$ is an intermediate variable, which can be expressed as:
\begin{equation}
    a = \frac{M}{2}\lambda_g.
\end{equation}

\noindent\textit{Proof.}
From (\ref{vM-v1}), the expected antenna array aperture $\mathbb{E}[x_M-x_1]$ can be readily obtained as:
\begin{equation}
    \mathbb{E}[x_M-x_1]=\mathbb{E}[v_M-v_1]=(M-1)\lambda_g.
\end{equation}
Due to the limited offset deviation, we approximate the PAs as being uniformly spaced. Under this assumption, the offset of the $m$-th PA, denoted by $\hat{v}_m$, can be expressed as follows:
\begin{equation}
\label{hat_vm}
    \hat{v}_m = \Big(m-\frac{M+1}{2}\Big)\lambda_g.
\end{equation}

Let $f_n(m)$ denote the array gain component contributed by the $m$-th PA when serving user $n$, i.e.,
\begin{equation}
    f_n(m)=\frac{1}{\sqrt{(x_c-x_n+\hat{v}_m)^2+y_n^2+h^2}}.
\end{equation}
By invoking the midpoint approximation, the discrete summation in (\ref{ArrayGainSum}) can be approximated by the integral:
\begin{equation}
\label{integralArrayGain}
\begin{split}
    \varepsilon_n 
    &= \sum_{m=1}^M 
    f_n(m) \approx 
    \int_{1/2}^{M+1/2} 
    f_n(m)dm.
\end{split}
\end{equation}
Based on (\ref{hat_vm}), by introducing the change of variable $v = \left(m - \frac{M+1}{2}\right)\lambda_g$, the integral can be reformulated as follows
\begin{equation}
\label{integral_epslon}
\begin{split}
    \varepsilon_n 
    &\approx 
    \int_{1/2}^{M+1/2} 
    f_n(m)dm.\\
    &= \frac{1}{\lambda_g}\int_{-a}^a\hat{f_n}(v)dv\\
    &= \frac{1}{\lambda_g}\int_{-a}^a\frac{1}{\sqrt{(x_c-x_n+v)^2+y_n^2+h^2}}dv\\
    &= \frac{1}{\lambda_g}\ln\Big(\sqrt{(x_c-x_n+v)^2+y_n^2+h^2}+x_c-x_n+v\Big)\Bigg|^a_{-a}.
\end{split}
\end{equation}
Then, by substituting the upper and lower bound into (\ref{integral_epslon}), the results in (\ref{ArrayGainResult}) can be obtained, which completes the proof.

Although (\ref{ArrayGainResult}) is exact under the integral approximation, its form is mathematically involved. To obtain more transparent insights, we further derive a simplified approximation.

\noindent\textbf{Corollary 1.} 
When $M$ PAs are optimally clustered and the user–array-center distance is sufficiently large compared with the array aperture, the effective array gain can be approximated as
\begin{equation}
\label{Corollary1}
    \varepsilon_n \approx \frac{M}{d_{n,c}}.
\end{equation}

\noindent\textit{Proof.}
Assuming that the expectation of array aperture $(M-1)\lambda_g$ is much smaller than the user–array-center distance $d_{n,c}$. We perform a second-order Taylor expansion of the integrand in (\ref{integral_epslon}) around $\mathbf{Q}_c$. Define
\begin{equation}
    \hat{f_n}(v)=\frac{1}{\sqrt{(x_c-x_n+v)^2+y_n^2+h^2}}.
\end{equation}
Its first- and second-order derivatives are
\begin{equation}
    \hat{f_n}'(v)=-\frac{x_c-x_n+v}{\big((x_c-x_n+v)^2+y_n^2+h^2\big)^{\frac{3}{2}}},
\end{equation}
\begin{equation}
    \hat{f_n}''(v)=\frac{2(x_c-x_n+v)^2-(y_n^2+h^2)}
    {\big((x_c-x_n+v)^2+y_n^2+h^2\big)^{\frac{5}{2}}}.
\end{equation}
Evaluating at $v=0$ yields
\begin{equation}
\begin{split}
    \hat{f_n}(0) & =\frac{1}{d_{n,c}},\\
    \hat{f_n}'(0) & =-\frac{x_c-x_n}{d_{n,c}^3},\\
    \hat{f_n}''(0) & =\frac{2(x_c-x_n)^2-(y_n^2+h^2)}{d_{n,c}^5}.    
\end{split}
\end{equation}

Neglecting higher-order terms, the integral in (\ref{integral_epslon}) can be approximated as (\ref{approxIntegral}).
\begin{figure*}
\begin{equation}
\label{approxIntegral}
    \varepsilon_n 
    \approx \frac{1}{\lambda_g}
    \int_{-a}^{a} 
    \left(
    \frac{1}{d_{n,c}}
    -\frac{x_c-x_n}{d_{n,c}^3}v
    +\frac{2(x_c-x_n)^2-(y_n^2+h^2)}{2d_{n,c}^5}v^2
    \right)\mathrm{d}v
    = \frac{M}{d_{n,c}}
    + \frac{M^3\lambda_g^2}{24}
    \frac{2(x_c-x_n)^2-(y_n^2+h^2)}{d_{n,c}^5}.
\end{equation}
\end{figure*}

Since $(M-1)\lambda_g \ll d_{n,c}$, the second term in (\ref{approxIntegral}) is negligible, which directly leads to (\ref{Corollary1}). This completes the proof.

Under the adopted TDMA protocol, the achievable rate of user $n$ can be characterized based on \textbf{Corollary~1} as follows.

\noindent\textbf{Lemma 2.}
Under the approximate array gain in (\ref{Corollary1}), the achievable rate of user $n$ is given by
\begin{equation}
\label{lemma2}
    R_n = \frac{1}{N}
    \log_2\left(
    1+\frac{M\gamma P}{d_{n,c}^2\sigma^2}
    \right).
\end{equation}

\section{Achievable Rate Optimization}

In this section, we investigate the structural properties of the multi-user sum-rate optimization problem. In particular, although the achievable rate of an individual user is generally non-concave, the aggregate sum rate exhibits a more favorable structure, being strictly concave in expectation and typically unimodal in practical finite-user scenarios, which enables efficient one-dimensional optimization. Based on suitable approximations, we further derive a closed-form approximate solution with low computational complexity. Finally, a geometric interpretation is provided to further reveal the structure of the theoretically optimal solution.

\subsection{Sum Rate Maximization Analysis}

It is observed from \eqref{x_m} and \eqref{v_m} that, once the array-center position $x_c$ and user location $\mathbf{U}_n$ are given, the corresponding offsets $\{v_m(t_n)\}_{m=1}^M$ can be determined, and hence the PA positions $\{\mathbf{Q}_m\}_{m=1}^M$ are uniquely specified. Therefore, the original sum-rate maximization problem in \eqref{probSumRate}, which is formulated with respect to all PA positions, can be reduced to a one-dimensional optimization problem over the array center only.

Specifically, since the waveguide is deployed along the $x$-axis and $\mathbf{Q}_c=[x_c,0,h]\in\mathcal{L}$ is fully determined by the scalar $x_c$, \eqref{probSumRate} can be rewritten as
\begin{subequations}\label{probSumRate_xc}
\begin{align}
\max_{x_c} \quad
& R_{\mathrm{sum}}(x_c)
= \sum_{n=1}^{N} R_n(x_c), \label{9a}\\
\text{s.t.} \quad
& 0 \le x_c \le D_1. \label{9b}
\end{align}
\end{subequations}

The reformulation in \eqref{probSumRate_xc} significantly simplifies the original optimization problem, since the optimization variables are reduced from the $M$ PA positions to the one-dimensional center position of the C-PAA. Based on this transformation, the subsequent analysis will focus on characterizing the achievable rate and the structural properties of $R_{\mathrm{sum}}(x_c)$.

Let $x_c=D_1r$ and $u_n = x_c-x_n$. Accordingly, $d_{n,c}^2 = u_n^2+y_n^2+h^2$. Taking the derivative of the achievable rate of user $n$ in (\ref{lemma2}) with respect to $u_n$, we obtain
\begin{equation}
\label{dRn_dun}
    \frac{\mathrm{d}R_{n}}{\mathrm{d}u_n} = -\frac{2\alpha}{N \ln2} \frac{u_n}{d_{n,c}^2 (d_{n,c}^2+\alpha)},
\end{equation}
where $\alpha$ is the performance coefficient, which can be expressed as
\begin{equation}
    \alpha = \frac{M\gamma P}{\sigma^2}.
\end{equation}

\noindent\textbf{Corollary 2.} 
When $|u_n| < \hat{D_n}$, the achievable rate of user $n$ is strictly concave. If this condition holds for all $N$ users, the resulting sum rate is also strictly concave. The bound parameter $\hat{D_n}$ can be expressed as follows
\begin{equation}
\label{Orig_D_n}
    \hat{D_n}=\sqrt{\frac{-(2\beta_n+\alpha) + \sqrt{16\beta_n^2+16\alpha \beta_n +\alpha^2}}{6}},
\end{equation}
where $\beta_n = y_n^2+h^2$.

\noindent\textit{Proof.}
Taking the second derivative of $R_n$, we obtain
\begin{equation}
\label{second_order_Rn_un}
    \frac{\mathrm{d}^2R_{n}}{\mathrm{d}u_n^2} = \frac{2\alpha}{N\ln2}\frac{3u_n^4+(2\beta_n+\alpha)u_n^2-\beta_n(\beta_n+\alpha)}{d_{n,c}^4(d_{n,c}^2+\alpha)^2}.
\end{equation}
Since strict concavity is equivalent to the second derivative being negative, the problem can be reformulated as follows:
\begin{equation}
\label{SecDerLess0un}
    3u_n^4+(2\beta_n+\alpha)u_n^2-\beta_n(\beta_n+\alpha)<0.
\end{equation}
Let $t=u_n^2$. To determine the critical point, (\ref{SecDerLess0un}) can be converted into the following equality:
\begin{equation}
    3t^2+(2\beta_n+\alpha)t-\beta_n(\beta_n+\alpha)=0.
\end{equation}
Solving the above equation yields
\begin{equation}
    t_{1,2} = \frac{-(2\beta_n+\alpha)\pm \sqrt{16\beta_n^2+16\alpha \beta_n +\alpha^2}}{6}.
\end{equation}
Under the constraint $t \geq 0$, the condition satisfying (\ref{SecDerLess0un}) is given by
\begin{equation}
\label{un_l_Dn}
    |u_n| < \sqrt{\frac{-(2\beta_n+\alpha) + \sqrt{16\beta_n^2+16\alpha \beta_n +\alpha^2}}{6}} = \hat{D_n}.
\end{equation}
For the multi-user sum rate, it can be readily obtained that
\begin{equation}
    \frac{\mathrm{d}^2R_{n}}{\mathrm{d}r^2} = D_1^2\frac{\mathrm{d}^2R_{n}}{\mathrm{d}u_n^2}.
\end{equation}
Therefore, if all $N$ users satisfy $|u_n| < \hat{D_n}$, the second derivative of the sum-rate function becomes
\begin{equation}
    \frac{\mathrm{d}^2R_{sum}}{\mathrm{d}r^2} = D_1^2\sum_{n=1}^N\frac{\mathrm{d}^2R_{n}}{\mathrm{d}u_n^2} < 0.
\end{equation}
Hence, $R_{sum}$ is also strictly concave. This completes the proof.

Since it is difficult to directly extract meaningful insights from (\ref{Orig_D_n}), we instead attempt to derive a more intuitive boundary condition.

\noindent\textbf{Remark 2.}
When $|x_c-x_n|<\sqrt{\beta_n/3}$, the achievable rate of user $n$ is strictly concave.

\noindent\textit{Proof.}
Let the test boundary be defined as $\overline{D}_n = \sqrt{k\beta_n}$, where $k$ denotes a distance scaling factor.  By applying (\ref{un_l_Dn}) and simplifying the condition $\hat{D}_n > \overline{D}_n$, we obtain:
\begin{equation}
\label{beta_n_inequality}
    (k+1)(3k-1)\beta_n \leq (1-k)\alpha.
\end{equation}
When $k = 1/3$, inequality (\ref{beta_n_inequality}) holds universally, thereby the second-order derivative of the achievable rate of user $n$ is strictly negative, which completes the proof.

\noindent\textbf{Theorem 1.}
Under the approximate rate model in \eqref{lemma2}, suppose that the user locations are uniformly distributed over the room floor, i.e.,
$X_n \sim U[0,D_1]$, $Y_n \sim U[0,D_2]$,
where $X_n$ and $Y_n$ are independent. Define the expected sum rate as
\begin{equation}
\bar R(x_c)
\triangleq
\mathbb{E}_{X,Y}\!\left[
\log_2\!\left(
1+\frac{\alpha}{(x_c-X)^2+Y^2+h^2}
\right)
\right].
\label{eq:expected_sum_rate}
\end{equation}
Then, $\bar R(x_c)$ is strictly concave on $(0,D_1)$. Consequently, $\bar R(x_c)$ is strictly unimodal over $[0,D_1]$ and admits a unique maximizer given by
\begin{equation}
x_c^\star=\frac{D_1}{2}.
\label{eq:midpoint_opt}
\end{equation}

\noindent\textit{Proof.}
Please refer to Appendix A.

It is worth emphasizing that the achievable rate of an individual user is generally not globally concave over the entire waveguide domain. As shown by (32), the second-order derivative of the user rate may become positive when the C-PAA center is sufficiently far from that user along the waveguide direction. Therefore, global concavity cannot, in general, be guaranteed at the single-user level.

Nevertheless, under the uniform user-distribution assumption, Theorem 1 shows that the expected sum-rate function is strictly concave with respect to the C-PAA center position. This result reveals an important smoothing effect of spatial averaging: although each individual user may exhibit a non-concave rate profile, averaging over the user distribution yields an expected sum-rate function that is strictly concave and thus has a unique maximizer.

For practical finite-user systems, a rigorous deterministic proof of unimodality for arbitrary user realizations remains difficult. However, motivated by the strict concavity of the expected objective and further supported by the numerical results in Section V, the realized sum-rate function is typically observed to exhibit a dominant single-peak profile over the feasible waveguide region, and in many cases behaves nearly or even strictly concavely. An intuitive explanation is that, according to (32), the contribution of a distant user to the curvature of the sum rate rapidly diminishes as the user--array-center distance increases, such that the overall shape is mainly governed by users located in the vicinity of the current C-PAA center. This observation provides both theoretical and numerical support for applying one-dimensional iterative search methods to determine the optimal C-PAA position.

To determine the optimal C-PAA position under a given realization of user locations, we transform the derivative in (\ref{dRn_dun}), originally taken with respect to $u_n$, into an equivalent expression with respect to the position parameter $r$:
\begin{equation}
\label{dRn_dr}
    \frac{\mathrm{d}R_{n}}{\mathrm{d}r} = -\frac{2\alpha}{N \ln2} \frac{D_1^2r-D_1x_n}{d_{n,c}^2 (d_{n,c}^2+\alpha)}.
\end{equation}
The derivative of the system sum rate with respect to $r$ can be expressed as follows:
\begin{equation}
\label{dRsum_dr}
    \frac{\mathrm{d}R_{sum}}{\mathrm{d}r} = \sum_{n=1}^N\frac{\mathrm{d}R_{n}}{\mathrm{d}r}.
\end{equation}

By leveraging the unimodal structure of the sum-rate function, the optimal center position of the C-PAA $\mathbf{Q}_c^{\star}$ can be determined using \textit{Algorithm~\ref{alg:C-PAA-optimization}}. The effectiveness of this algorithm is further validated by the numerical results in Section~V.

\begin{algorithm}[h]
\caption{Optimal C-PAA Center Location Optimization}
\label{alg:C-PAA-optimization}
\begin{algorithmic}[1]
\REQUIRE Initial point $\mathbf{A}$, ending point $\mathbf{B}$, user positions $\{\mathbf{U}_n\}_{n=1}^N$, performance coefficient $\alpha$
\ENSURE Optimal C-PAA center $\mathbf{Q}_c^{\star}$

\STATE Define the derivative function:
\begin{equation*}
    g'(r) = \frac{\mathrm{d}R_{\mathrm{sum}}(r)}{\mathrm{d}r}
\end{equation*}

\STATE Evaluate $g'(0)$ and $g'(1)$

\IF{$g'(0)\cdot g'(1) \leq 0$}
    \STATE Find $r^{\star}$ such that $g'(r^{\star}) = 0$ using a root-finding method (e.g., bisection or \texttt{fzero}) over $[0,1]$
\ELSE
    \STATE Evaluate boundary values:
    \begin{equation*}
        R_0 = R_{\mathrm{sum}}(\mathbf{A}), \quad
        R_1 = R_{\mathrm{sum}}(\mathbf{B})
    \end{equation*}
    \STATE Select
    \begin{equation*}
        r^{\star} =
        \begin{cases}
            0, & \text{if } R_0 \geq R_1 \\
            1, & \text{otherwise}
        \end{cases}
    \end{equation*}
\ENDIF

\STATE Update the optimal C-PAA center:
\begin{equation*}
    \mathbf{Q}_c^{\star} = \mathbf{A} + (\mathbf{B}-\mathbf{A})r^{\star}
\end{equation*}

\RETURN $\mathbf{Q}_c^{\star}$
\end{algorithmic}
\end{algorithm}

The computational complexity of Algorithm~\ref{alg:C-PAA-optimization} primarily depends on the number of users $N$ and the presence of a stationary point along the search direction. Specifically, if a root $r^\star \in (0,1)$ exists, a one-dimensional root-finding procedure is used to solve $f'(r) = 0$. Assuming a target accuracy of $\epsilon$, the bisection method requires $O(\log(1/\epsilon))$ iterations, and each iteration involves evaluating the derivative of the sum-rate with respect to $r$, which has a complexity of $O(N)$. Therefore, in this case, the overall computational complexity is $O(N \log(1/\epsilon))$.  

In the absence of a root within $(0,1)$, the algorithm simply evaluates the sum-rate at the boundaries $r=0$ and $r=1$, resulting in an overall complexity of $O(N)$.  

\subsection{Closed-Form Approximate Optimal Solution}

To further reduce computational complexity, in this subsection, we leverage expectation-based analysis to derive a closed-form approximate solution for the optimal C-PAA position.

The sum-rate maximization problem in (\ref{probSumRate_xc}) can be interpreted as a logarithmic variant of the one-dimensional \textit{Weber–Fermat} problem \cite{Chandrasekaran1990FermatWeber}. As it does not admit an exact closed-form solution, approximation methods are therefore required. To fully exploit the user location information and obtain a tractable solution, the derivative of the sum rate is approximated as in (\ref{dRsumExpectation}), where $\mathbb{E}_{x_c}[\cdot]$ and $\mathbb{E}_{x_c,x_n,y_n}[\cdot]$ denote the expectations taken with respect to the distributions of $x_c$, and the joint distributions of $x_c$, $x_n$, and $y_n$, respectively.
Under the same uniform distribution assumption, we model $x_c$, $x_n$, and $y_n$ as independent random variables with $x_c,x_n\sim U[0,D_1]$ and $y_n\sim U[0,D_2]$.

\begin{figure*}
\begin{equation}
\label{dRsumExpectation}
\begin{split}
    \frac{\mathrm{d}R_{sum}}{\mathrm{d}r} &= \sum_{n=1}^N\Big(-\frac{2\alpha}{N\ln2} \frac{D_1^2r-D_1x_n}{d_{n,c}^2 (d_{n,c}^2+\alpha)}\Big) 
    \approx \frac{2\alpha D_1}{N\ln 2}\sum_{n=1}^N \Big(\frac{x_n}{\mathbb{E}_{x_c}[d_{n,c}^2 (d_{n,c}^2+\alpha)]}-\frac{D_1r}{\mathbb{E}_{x_c,x_n,y_n}[d_{n,c}^2 (d_{n,c}^2+\alpha)]}\Big)
\end{split}
\end{equation}  
\end{figure*}

Exploiting the unimodal behavior of the system sum rate, we approximate the stationary condition by setting (\ref{dRsumExpectation}) to zero. After algebraic manipulations, the approximate optimal array position parameter is obtained as

\begin{equation}
\label{r_star}
\begin{split}
    r^{\star}_{approx}=&\frac{\mathbb{E}_{x_c,x_n,y_n}[d_{n,c}^2(d_{n,c}^2+\alpha)]}{ND_1}\\
    &\times\sum_{n=1}^N\frac{x_n}{\mathbb{E}_{x_c}[d_{n,c}^2(d_{n,c}^2+\alpha)]}. 
\end{split}   
\end{equation}

If $r^{\star}_{approx}$ lies outside the feasible region, it is projected onto the nearest boundary point, i.e., $0$ or $1$.
The expectation terms in (\ref{r_star}) admit closed-form expressions, as summarized in the following proposition.

\noindent\textbf{Proposition 2.}
Under the uniform distribution assumptions, $\mathbb{E}_{x_c,x_n,y_n}[d_{n,c}^2(d_{n,c}^2+\alpha)]$ and $\mathbb{E}_{x_c}[d_{n,c}^2(d_{n,c}^2+\alpha)]$ are given by
\begin{equation}
\begin{split}
\mathbb{E}_{x_n,x_c,y_n}[d_{n,c}^2(d_{n,c}^2+\alpha)]
=&
\frac{D_1^4}{15}
+\frac{D_2^4}{5}
+\frac{D_1^2D_2^2}{9}\\
&+\frac{h^2D_1^2}{3}
+\frac{2h^2D_2^2}{3}
+h^4\\
&+\alpha\left(\frac{D_1^2}{6}+\frac{D_2^2}{3}+h^2\right),
\end{split}
\end{equation}
and
\begin{equation}
\begin{split}
\mathbb{E}_{x_c}[d_{n,c}^2(d_{n,c}^2+\alpha)]
=&\delta_n^4+\left(\frac{D_1^2}{2}+2\beta_n+\alpha\right)\delta_n^2
+\frac{D_1^4}{80}\\
&+\frac{(2\beta_n+\alpha)D_1^2}{12}
+\beta_n(\beta_n+\alpha),
\end{split}
\end{equation}
where $\delta_n = x_n - \frac{D_1}{2}$ and $\beta_n = y_n^2 + h^2$.

\noindent\textit{Proof.}
Please refer to Appendix B.

For the special case $D_1 = D_2 = D$, the first expectation simplifies to
\begin{equation}
\begin{split}
\mathbb{E}_{x_n,x_c,y_n}[d_{n,c}^2(d_{n,c}^2+\alpha)]
=&
\frac{17}{45}D^4+D^2h^2+h^4\\
&+\alpha\left(\frac{D^2}{2}+h^2\right).
\end{split}
\end{equation}

Substituting the above results into (\ref{r_star}) yields a fully explicit expression for the approximate optimal position parameter $r^{\star}_{\mathrm{approx}}$, and its effectiveness is further verified by the numerical results in Section~V.

\subsection{Geometric Interpretation}

To gain further insight into the structure of the optimal solution, a geometric interpretation of the sum-rate maximization problem is provided. Specifically, it is observed that the first-order derivative of the user rate in (\ref{dRn_dr}) admits a vector inner-product form. Building upon (\ref{dRsum_dr}), the derivative of the sum rate can be rewritten as:
\begin{equation}
\begin{split}
    \frac{\mathrm{d}R_{sum}}{\mathrm{d}r} = -\frac{2\alpha}{N \ln 2}\sum_{n=1}^N \frac{(\mathbf{Q}_c-\mathbf{U}_n)\cdot \mathbf{d}}{d_{n,c}^2 (d_{n,c}^2+\alpha)},
\end{split}
\end{equation}
where $\mathbf{d}$ denotes the direction vector of the waveguide, i.e.,
\begin{equation}
    \mathbf{d} = \mathbf{B}-\mathbf{A},
\end{equation}
and $\cdot$ denotes the Euclidean inner product between two vectors.
The inner product reveals that the gradient of the sum rate is determined by the projection of the user-to-array-center vectors onto the waveguide direction. Consequently, the optimality condition for maximizing the sum rate can be equivalently characterized by an orthogonality condition, i.e.,
\begin{equation}
\label{perpd}
    \frac{\mathrm{d}R_{sum}}{\mathrm{d}r} = 0 \Rightarrow \sum_{n=1}^N\frac{\mathbf{U}_n-\mathbf{Q}_c}{d_{n,c}^2 (d_{n,c}^2+\alpha)} \perp \mathbf{d},
\end{equation}
where $\perp$ denotes vector orthogonality.

The above condition admits an intuitive geometric interpretation: at the optimum, the weighted sum of the user-to-center vectors has zero projection onto the waveguide direction $\mathbf{d}$.

\noindent\textbf{Remark 3.}
When the length of the waveguide matches the room dimension along the same direction, the optimal array center $\mathbf{Q}_c^{\star}$  lies strictly in the interior of the waveguide, rather than at its boundary.

\noindent\textit{Proof.}
Without loss of generality, assume that $\mathbf{d}$ is aligned with the $x$-axis. Then, the orthogonality condition in (\ref{perpd}) reduces to:
\begin{equation}
    \sum_{n=1}^N \frac{x_n - x_c}{d_{n,c}^2 (d_{n,c}^2+\alpha)} = 0.
\end{equation}
Since all weights $d_{n,c}^2 (d_{n,c}^2+\alpha)$ are strictly positive, the above equation implies that $x_c$ must lie within the convex hull of $\{x_n\}_{n=1}^N$, i.e., between the minimum and maximum user locations along the $x$-axis.

When the waveguide spans the entire room dimension, this interval is fully contained within the feasible region of the waveguide, thereby ensuring that the optimal solution lies in its interior. In contrast, if the waveguide is shorter than the room dimension, the feasible region may truncate this interval, in which case the optimum can occur at the boundary.

\section{Numerical Results}

\begin{figure*}[!t]
\centering
\subfloat[Effective array gain versus $x_c$.]{\label{gainVsX_c}\includegraphics[width=2.35in]{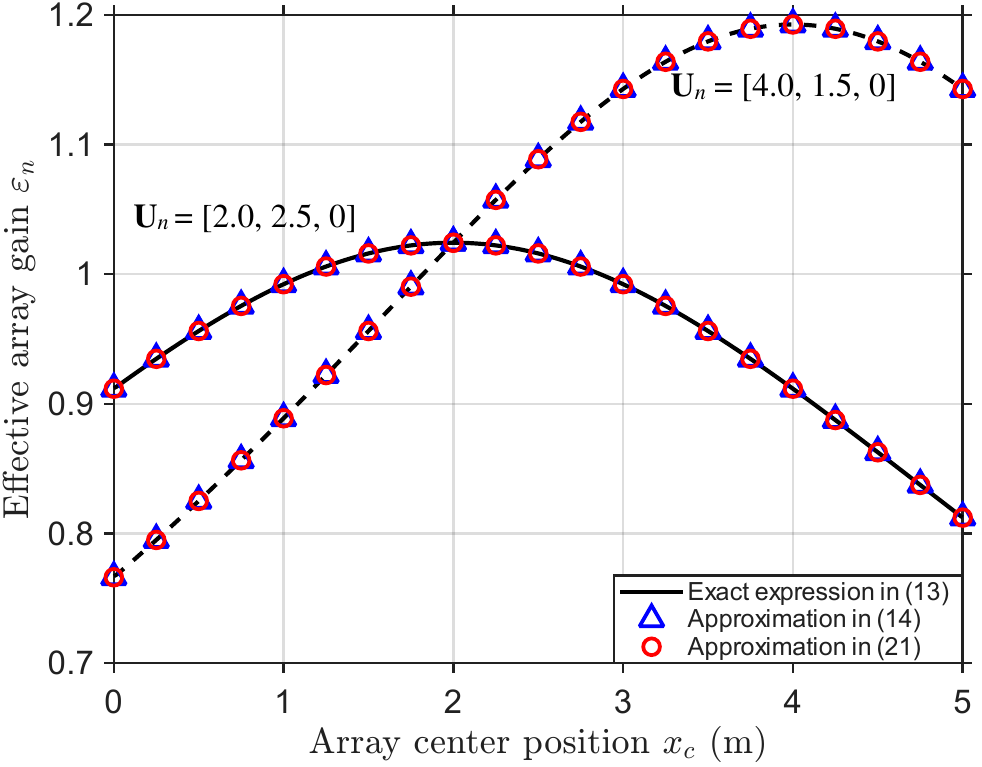}}
\subfloat[Relative error of $\varepsilon_n$ versus $M$.]{\label{RelativeErrorVsM}\includegraphics[width=2.35in]{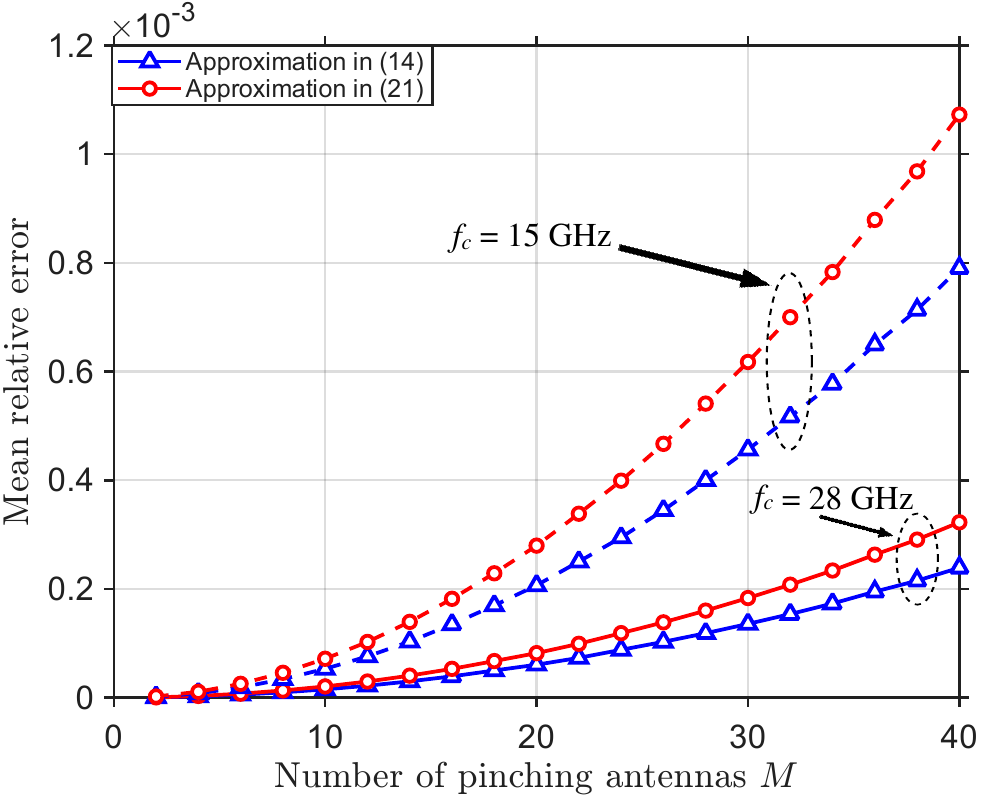}}
\subfloat[Sum-rate error versus $P$.]{\label{RateErrorVsP}\includegraphics[width=2.35in]{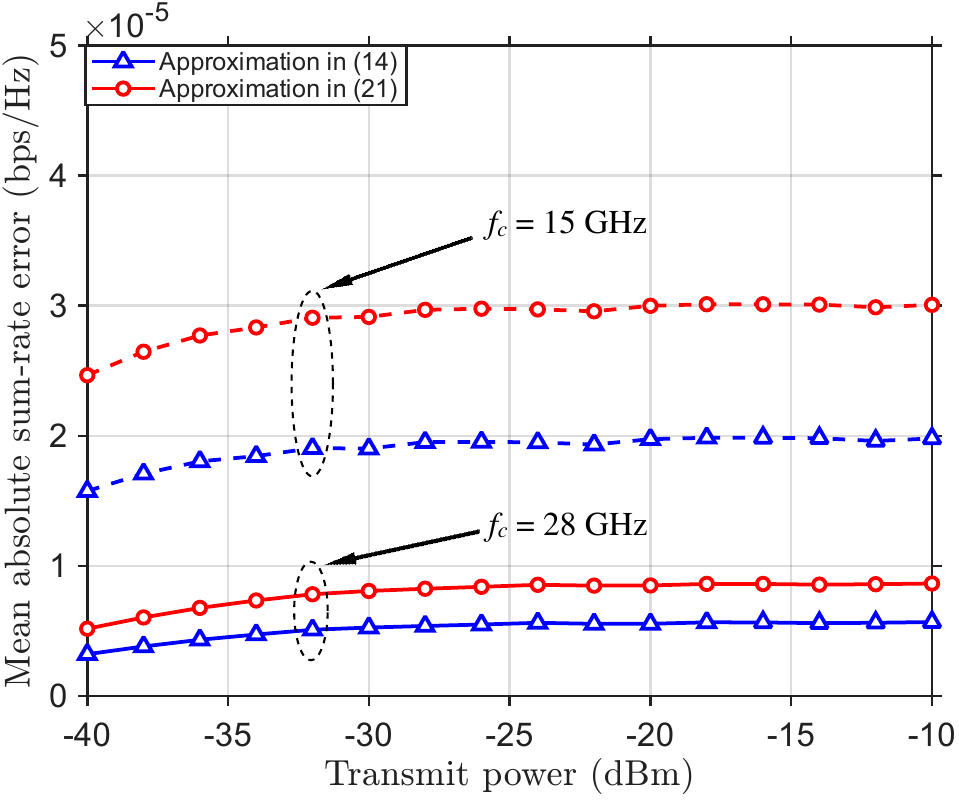}}
\caption{Approximation verification of effective channel gain.}
\label{fig:EffectiveChannelGain}
\end{figure*}

In this section, numerical results are presented to evaluate the performance of the proposed C-PAA system. Monte Carlo simulations are conducted to verify the accuracy of the proposed analytical results.

Unless otherwise specified, we consider an indoor room of size $D_1 \times D_2 \times h = 5 \times 5 \times 3~\mathrm{m}^3$, with $N=8$ users randomly and uniformly distributed over the floor and $M=4$ PAs.
The carrier frequency is set to $f_c = 28~\mathrm{GHz}$, and the transmission bandwidth is $BW = 10~\mathrm{MHz}$. The AWGN power is determined by the bandwidth and modeled as $\sigma^2 = -174 + 10\log_{10}(BW)$ dBm. The transmit power is set to $-20$ dBm, and the effective refractive index of the waveguide is chosen as $n_{\mathrm{eff}} = 1.4$~\cite{pozar2021microwave}.

\subsection{Validation of Theoretical Analysis}
To validate the accuracy of the theoretical derivations and approximations presented in this paper, this subsection employs numerical simulations to verify the approximated effective channel gain, the unimodal property of the sum rate, and the accuracy of the derived approximate optimal array position.
\subsubsection{Effective Channel Gain}

To verify the accuracy of the approximations developed in \textbf{Lemma 1} and \textbf{Corollary 1}, Fig.~\ref{gainVsX_c} compares the approximate effective array gain $\varepsilon_n$ given by \eqref{ArrayGainResult} and \eqref{Corollary1} with the exact result in \eqref{ArrayGainSum} as a function of the array-center position $x_c$, for two representative user locations $[2,\,2.5,\,0]$ and $[4,\,1.5,\,0]$. It is observed that both approximations closely match the exact result over the entire range of $x_c$, thereby confirming their high accuracy.

Fig.~\ref{RelativeErrorVsM} shows the average relative error of the array gain as a function of the number of PAs, where the array-center positions and user locations are randomly generated over $10^6$ Monte Carlo realizations. As expected, the approximation in \eqref{Corollary1} yields a slightly larger relative error than that in \eqref{ArrayGainResult}, since it is obtained through a further simplification of \eqref{ArrayGainResult}. In addition, the relative error gradually increases with the number of antennas. This is because that enlarging the array aperture strengthens the effect of the higher-order terms omitted in the Taylor expansion in \eqref{approxIntegral}, which leads to a larger approximation mismatch. Nevertheless, the overall error remains very small even for relatively large antenna numbers, owing to the high carrier frequency considered in the PASS setup, which confirms the practical accuracy of both approximations.

Fig.~\ref{RateErrorVsP} further evaluates the analytical approximation error in terms of the system sum rate under different transmit power levels. It can be seen that the approximation based on \eqref{ArrayGainResult} consistently achieves a smaller error than that based on \eqref{Corollary1}. Consistent with the observation in Fig.~\ref{RelativeErrorVsM}, the error at $15$~GHz remains larger than that at $28$~GHz, since a lower carrier frequency leads to a larger array aperture for the same number of PAs and therefore increases the approximation mismatch. Moreover, both approximations exhibit only weak sensitivity to the transmit power and maintain a low error level throughout the considered power range. These results further demonstrate the robustness and practical usefulness of the proposed analytical approximations.

\subsubsection{Unimodal Property of the Sum-rate}



\begin{figure}[b]
    \centering
    \includegraphics[width=0.85\linewidth]{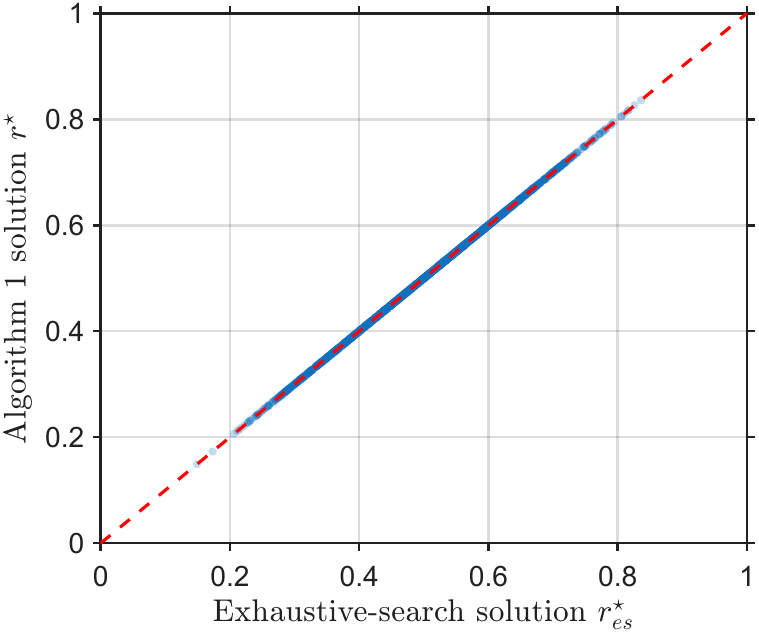}
    \caption{Scatter plot of Algorithm 1 versus exhaustive search.}
    \label{alg1VsExhaus}
\end{figure}

The unimodal structure of the sum-rate function is critical for solving the optimization problem via Algorithm~1. To verify the unimodal property of the system sum rate, the solution obtained by Algorithm~1 is compared with the benchmark solution obtained by exhaustive search along the waveguide under randomly generated user distributions, as illustrated in the scatter plot of Fig.~\ref{alg1VsExhaus}. It can be seen that the optimal array positions returned by Algorithm~1 coincide exactly with those obtained by exhaustive search. This result not only validates the unimodal behavior of the sum-rate function but also confirms the accuracy of the proposed iterative optimization algorithm.

\begin{figure*}[b]
\centering
\subfloat[Approximation versus exact solution.]{\label{rappVsrstar}\includegraphics[width=2.35in]{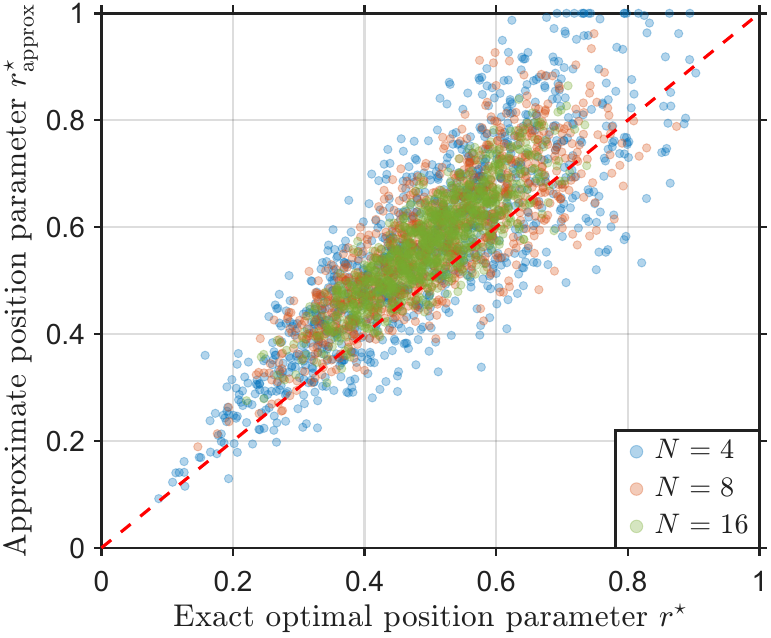}}
\subfloat[CDF of approximate error.]{\label{cdf_rapp_error}\includegraphics[width=2.35in]{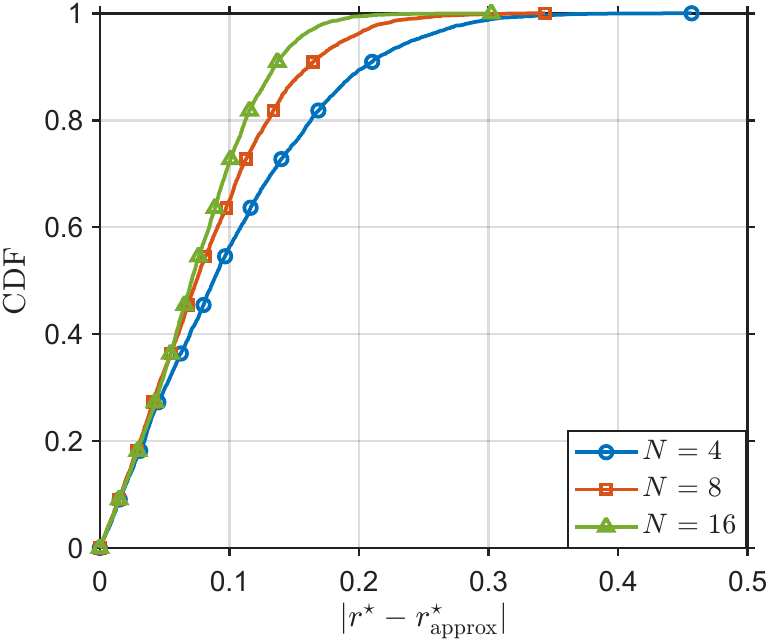}}
\subfloat[Sum rate error versus $P$]{\label{sumRate_gap_vs_power}\includegraphics[width=2.35in]{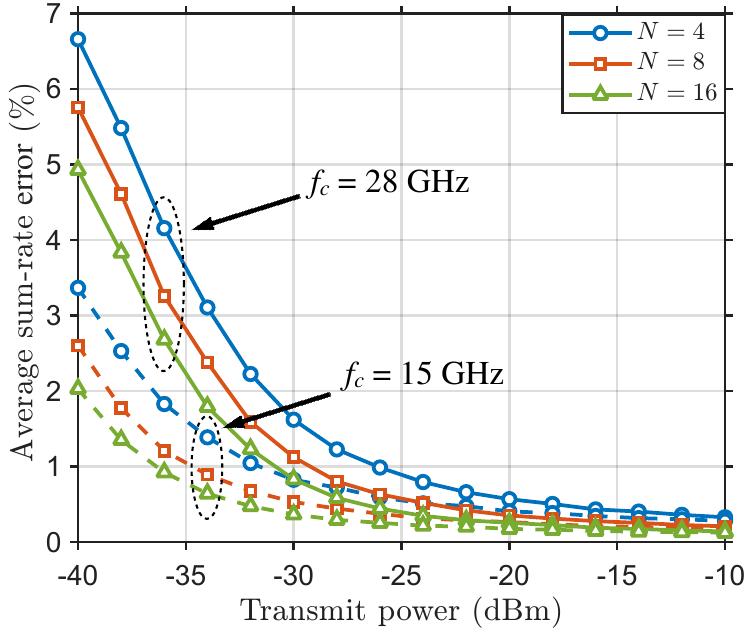}}
\caption{Verification of closed-form approximate optimal array position}
\label{fig:cf_approxmation}
\end{figure*}

\subsubsection{Approximate Optimal Array Position}

Although Algorithm~1 provides a low-complexity method for determining the optimal array position, it still relies on iterative computation. To further reduce the computational complexity, a closed-form approximate solution for the optimal array position was derived in \eqref{r_star}, and its accuracy is evaluated in Fig.~\ref{fig:cf_approxmation}.

Fig.~\ref{rappVsrstar} presents a scatter plot comparing the approximate solution in \eqref{r_star} with the exact solution obtained via Algorithm~1 under randomly distributed user locations. It is observed that the approximate solutions are tightly clustered around the exact ones. Fig.~\ref{cdf_rapp_error} further shows the cumulative distribution function (CDF) of the approximation error for different numbers of users, which further confirms the accuracy of the proposed approximation. Overall, these results indicate that the approximation becomes more accurate as the number of users increases, which is consistent with its expectation-based derivation.

Fig.~\ref{sumRate_gap_vs_power} illustrates the percentage error in the average sum rate between the exact and approximate solutions versus the transmit power under different carrier frequencies. It can be observed that the average sum-rate error decreases as the transmit power increases for all considered carrier frequencies.
As the transmit power increases, the sum rate becomes less sensitive to array-position mismatch, and the percentage gap between the exact and approximate solutions correspondingly decreases. Since \eqref{r_star} is obtained from an expectation-based approximation, its accuracy improves with the number of users, leading to a smaller gap in larger-user scenarios. A slightly larger gap is also observed at higher carrier frequencies, where more severe path loss increases the sensitivity of the achievable rate to approximation errors in the array position.

\subsection{Performance Optimization Comparison}
To evaluate the effectiveness of the proposed C-PAA, the optimized system performance is compared with the theoretical upper bound and existing pinching antenna array schemes, followed by a comprehensive performance analysis.

\subsubsection{Performance Upper Bound}
\begin{figure}[b]
    \centering
    \includegraphics[width=0.9\linewidth]{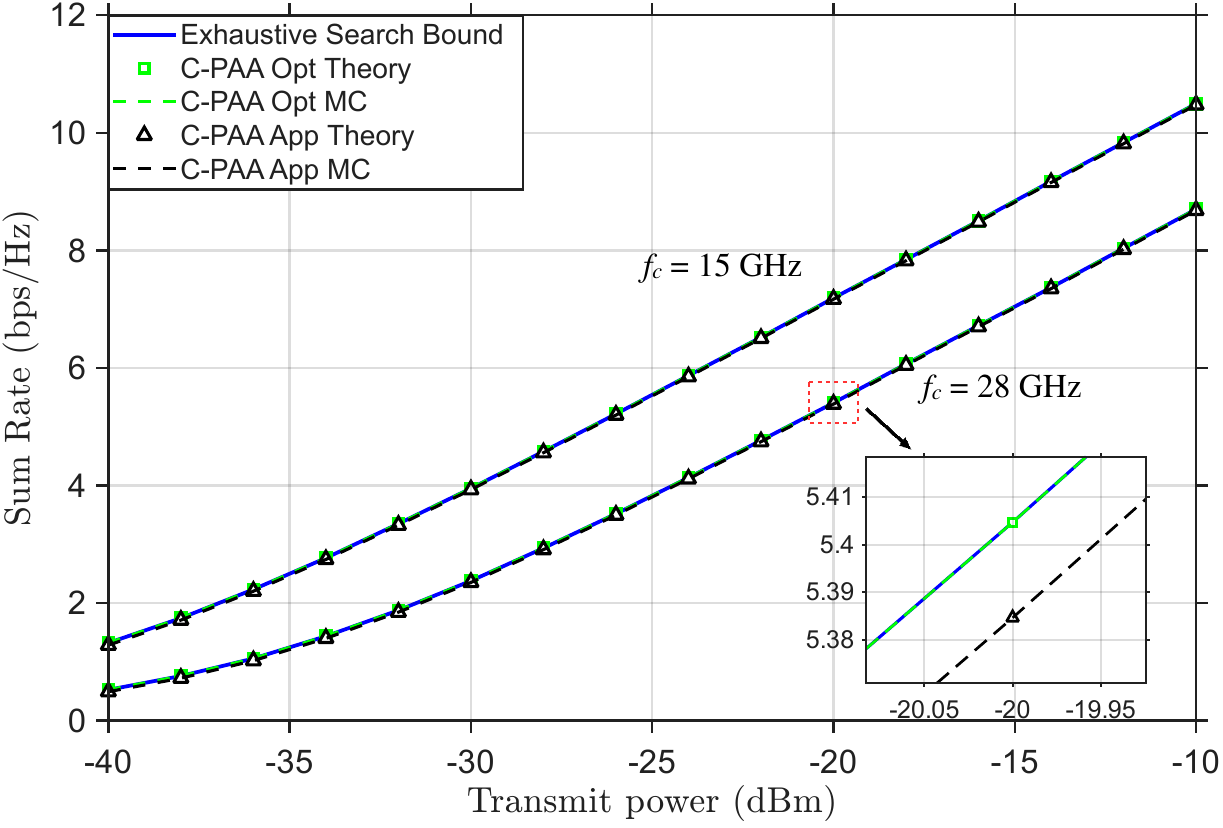}
    \caption{Upper bound analysis of sum rate.}
    \label{fig:UpperBound}
\end{figure}

To characterize the performance upper bound of the proposed C-PAA, an ideal benchmark is constructed by considering the same number of PAs while neglecting both the mutual coupling among antennas and the physical constraints imposed by the pinching devices. In addition, the phase shifts introduced by electromagnetic propagation in both the waveguide and free space are ignored, such that the signals transmitted from all PAs are assumed to be perfectly phase-aligned at the user side. Under this idealized setting, the theoretical upper bound on the achievable rate is obtained via exhaustive search. The corresponding performance comparison between this upper bound and the proposed optimization scheme is presented in Fig.~\ref{fig:UpperBound}.

As shown in Fig.~\ref{fig:UpperBound}, under different carrier frequencies and transmit power levels, the sum-rate performance achieved by the proposed C-PAA closely approaches the ideal upper bound, with only a marginal gap. The closed-form approximate solution also remains close to these two curves. The small gap of curves confirms the high accuracy of the proposed approximation and the effectiveness of the proposed C-PAA design in TDMA scenarios.
A closer examination indicates that the optimal antenna configuration obtained by exhaustive search corresponds to the case where all antennas are effectively co-located at the same position. Notably, this position is highly consistent with the solution returned by Algorithm~1, which explains why the proposed optimization scheme can achieve near-upper-bound performance.
Furthermore, the theoretical results corresponding to both the iterative optimal solution and the closed-form approximate solution are in close agreement with their respective Monte Carlo simulation results, thereby further confirming the accuracy of the analytical framework developed in this paper.

\subsubsection{Performance Comparison with Benchmarks}

\begin{figure*}[b]
\centering
\subfloat[Sum rate versus user number $N$.]{\label{SRvsN}\includegraphics[width=2.35in]{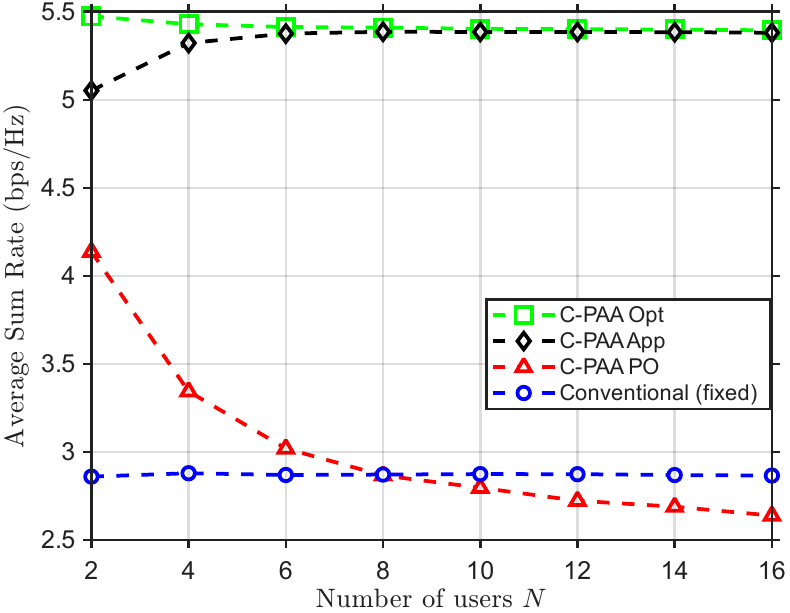}}
\subfloat[Sum rate versus PA number $M$.]{\label{SRvsM}\includegraphics[width=2.35in]{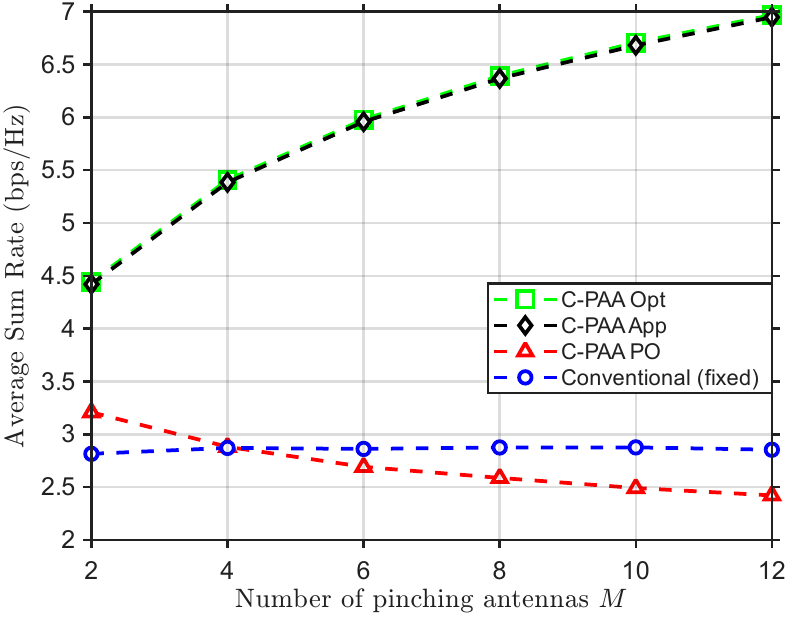}}
\subfloat[Sum rate versus transmit power $P$.]{\label{SRvsTP}\includegraphics[width=2.35in]{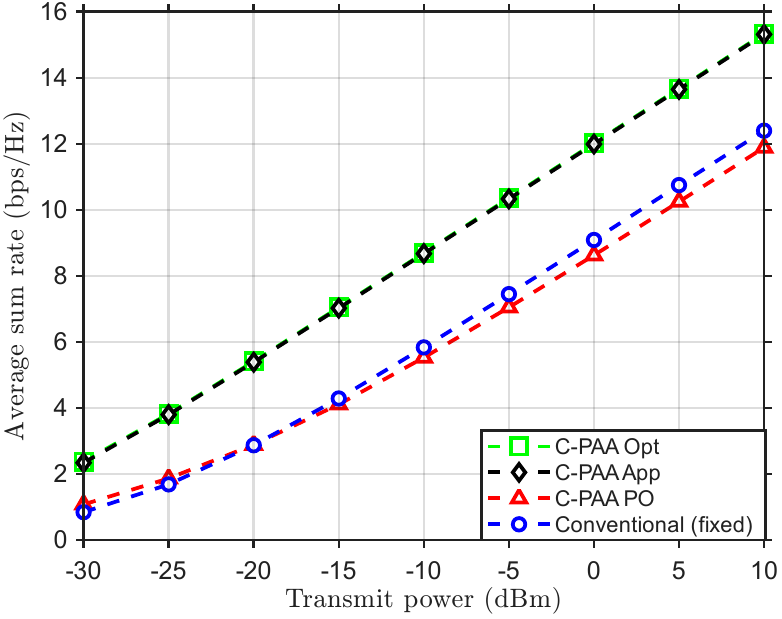}}
\caption{Average sum rate comparison with benchmark schemes.}
\label{fig:SRvsBenchmark}
\end{figure*}

To further evaluate the effectiveness of the proposed C-PAA, we compare it with two benchmark single-waveguide multi-PA TDMA schemes, namely, the conventional fixed-spacing PA array and the priority-oriented (PO) C-PAA scheme proposed in \cite{houPerformancePAA}. In the conventional scheme, the PAs are uniformly distributed along the waveguide with fixed spacing, whereas in the PO scheme, the array is steered toward the user closest to the waveguide.

Fig. \ref{fig:SRvsBenchmark} presents the average sum-rate performance under randomly distributed user locations from three perspectives, namely, versus the number of users $N$, the number of PAs $M$, and the transmit power $P$. The three subfigures consistently show that the proposed C-PAA scheme outperforms the benchmark schemes in terms of system sum rate.

As shown in Fig. \ref{SRvsN}, the proposed C-PAA optimization scheme consistently achieves the highest sum rate over the entire range of user numbers, while the closed-form approximation remains very close to the optimal solution. Moreover, as the number of users increases, the performance of the approximate solution progressively approaches that of the iterative optimal solution. In contrast, the PO scheme exhibits a clear performance degradation as the number of users increases. This is because prioritizing a single user becomes increasingly suboptimal in dense-user scenarios, where the overall system performance is determined by the spatial balance among multiple users rather than by the channel quality of one favored user. The conventional fixed-spacing scheme is less sensitive to the number of users, but its performance remains consistently inferior due to the lack of adaptive array-position optimization.

Fig. \ref{SRvsM} illustrates the average sum rate as a function of the number of PAs. It can be observed that both the proposed C-PAA and its closed-form approximation benefit significantly as the number of PAs increases, and their achievable sum rates increase monotonically, which demonstrates that the proposed scheme can effectively exploit the beamforming and coherent combining gains provided by a larger number of pinching antennas.
By comparison, the performance of the PO scheme decreases as the number of PAs increases. The reason is that a larger number of PAs leads to a narrower beam, which favors the prioritized user but degrades the rates of the remaining users, thereby reducing the overall sum rate. 
Meanwhile, the conventional fixed-spacing array exhibits only marginal improvement, which further highlights the importance of adaptive C-PAA positioning.

Fig.~\ref{SRvsTP} further shows the average sum rate versus the transmit power $P$. It is observed that the sum rates achieved by all schemes increase with the increasing transmit power, while the proposed C-PAA consistently attains the best performance over the entire power range. Moreover, the approximate solution remains nearly indistinguishable from the iterative one, demonstrating its robustness under different SNR conditions. Overall, these results verify that the proposed C-PAA achieves substantial performance gains over the benchmark schemes while providing a near-optimal solution with low computational complexity for multi-user PASS optimization.

\section{Conclusion}

In this paper, we investigated the design and performance optimization of a constrained pinching antenna array for multi-user downlink PASS. A sum-rate maximization problem was formulated by jointly optimizing the array-center position and the wavelength-scale PA configuration within the array under practical array-size constraints. To address this problem, we first characterized the structural properties of the C-PAA and derived an explicit upper bound on its aperture. Then, tractable approximations for the effective channel gain and the achievable rate were developed. Building on these results, we analyzed the multi-user sum-rate optimization problem and showed that the system sum rate exhibits a unimodal structure under practically relevant conditions. To further reduce the computational complexity, a closed-form approximate solution for the near-optimal array-center position was also derived based on expectation analysis.

Numerical results verified the accuracy of the proposed analysis and demonstrated that the C-PAA can closely approach the ideal upper bound while consistently outperforming conventional fixed-spacing and priority-oriented benchmark schemes. Moreover, the closed-form approximation was shown to achieve performance close to that of the iterative optimum, especially in large-user scenarios. Overall, this work provided a low-complexity design framework and useful theoretical insights for constrained movable PA arrays in future indoor multi-user PASS deployments.

\appendices

\section{Proof of Theorem 1}

Under the approximate rate expression in \eqref{lemma2}, the expected sum rate under uniformly distributed user locations can be written as
\begin{equation}
\bar{R}(x_c)=\mathbb{E}_{X,Y}\!\left[\log_2\!\left(1+\frac{\alpha}{(x_c-X)^2+Y^2+h^2}\right)\right],
\end{equation}
where $X\sim U[0,D_1]$ and $Y\sim U[0,D_2]$ are independent random variables. For ease of exposition, define
\begin{equation}
R(x_c;X,Y)\triangleq \log_2\!\left(1+\frac{\alpha}{(x_c-X)^2+Y^2+h^2}\right).
\end{equation}
Hence, by differentiating with respect to $x_c$, we obtain 
\begin{equation}
\begin{aligned}
\frac{\mathrm{d} R(x_c;X,Y)}{\mathrm{d} x_c}
=
&-\frac{2\alpha(x_c-X)}
{\ln2\left((x_c-X)^2+\beta\right)} \\
&\times\frac{1}
{\left((x_c-X)^2+\beta+\alpha\right)},
\end{aligned}
\label{eq:dRdx_generic}
\end{equation}
where $\beta \triangleq Y^2+h^2$.

Taking expectation with respect to $(X,Y)$ yields 
\begin{equation}
\bar R'(x_c)
=
\mathbb{E}_{X,Y}
\!\left[
\frac{\mathrm{d} R(x_c;X,Y)}{\mathrm{d} x_c}
\right].
\label{eq:Rbar_prime_start}
\end{equation}
Since $X\sim U[0,D_1]$ and $Y\sim U[0,D_2]$, we can rewrite \eqref{eq:Rbar_prime_start} as \eqref{eq:Rbar_prime_integral}.
\begin{figure*}
\begin{equation}
\bar R'(x_c)
=
-\frac{2\alpha}{D_1D_2\ln 2}
\int_0^{D_2}\int_0^{D_1}
\frac{x_c-x}{\left((x_c-x)^2+\beta\right)\left((x_c-x)^2+\beta+\alpha\right)}
\,dx\,dy.
\label{eq:Rbar_prime_integral}
\end{equation}
\end{figure*}

For any fixed $y$, letting $u=x_c-x$ gives \eqref{eqint}.
\begin{figure*}
\begin{equation}
\label{eqint}
\int_0^{D_1}
\frac{x_c-x}{\left((x_c-x)^2+\beta\right)\left((x_c-x)^2+\beta+\alpha\right)}dx
=
\int_{x_c-D_1}^{x_c}
\frac{u}{(u^2+\beta)(u^2+\beta+\alpha)}\,du.
\end{equation}
\end{figure*}
Using 
\begin{equation}
    \int
\frac{u}{(u^2+\beta)(u^2+\beta+\alpha)}\,du
=
\frac{1}{2\alpha}
\ln\frac{u^2+\beta}{u^2+\beta+\alpha},
\end{equation}
we obtain
\begin{equation}
\bar R'(x_c)
=
-\frac{1}{D_1D_2\ln 2}
\int_0^{D_2}
\left[
\Psi_\beta(x_c)-\Psi_\beta(D_1-x_c)
\right]dy,
\label{eq:Rbar_prime_closed}
\end{equation}
where
\begin{equation}
\Psi_\beta(t)
\triangleq
\ln\!\left(\frac{t^2+\beta}{t^2+\beta+\alpha}\right),
\qquad t\ge 0.
\label{eq:Psi_def}
\end{equation}

Next, differentiating $\Psi_\beta(t)$ gives
\begin{equation}
\Psi_\beta'(t)
=
\frac{2\alpha t}{(t^2+\beta)(t^2+\beta+\alpha)} > 0,
\qquad t>0.
\label{eq:Psi_monotone}
\end{equation}
Therefore, $\Psi_\beta(t)$ is strictly increasing on $[0,\infty)$. It follows from \eqref{eq:Rbar_prime_closed} that
\begin{equation}
    \bar R'(x_c)
\begin{cases}
>0, & 0\le x_c < \frac{D_1}{2},\\[1mm]
=0, & x_c = \frac{D_1}{2},\\[1mm]
<0, & \frac{D_1}{2} < x_c \le D_1.
\end{cases}
\end{equation}
Hence, $\bar R(x_c)$ is strictly unimodal over $[0,D_1]$ and its unique maximizer is $x_c^\star=D_1/2$.

To establish strict concavity, differentiate \eqref{eq:Rbar_prime_closed} with respect to $x_c$. By the chain rule, we have
\begin{equation}
\bar R''(x_c)
=
-\frac{1}{D_1D_2\ln 2}
\int_0^{D_2}
\left[
\Psi_\beta'(x_c)+\Psi_\beta'(D_1-x_c)
\right]dy.
\label{eq:Rbar_second}
\end{equation}
Substituting \eqref{eq:Psi_monotone} into \eqref{eq:Rbar_second}, for any $x_c\in(0,D_1)$, we obtain \eqref{eq:Rbar_second_negative}.
\begin{figure*}
\begin{equation}
\bar R''(x_c)
=
-\frac{2\alpha}{D_1D_2\ln 2}
\int_0^{D_2}
\left[
\frac{x_c}{(x_c^2+\beta)(x_c^2+\beta+\alpha)}
+
\frac{D_1-x_c}{((D_1-x_c)^2+\beta)((D_1-x_c)^2+\beta+\alpha)}
\right]dy
<0.
\label{eq:Rbar_second_negative}
\end{equation}
\end{figure*}
Therefore, $\bar R(x_c)$ is strictly concave on $(0,D_1)$. The proof is completed.

\section{Proof of Proposition 2}

We first evaluate $\mathbb{E}_{x_c,x_n,y_n}[d_{n,c}^2(d_{n,c}^2+\alpha)]$. For notational brevity, we shall denote $\mathbb{E}_{x_c,x_n,y_n}[\cdot]$ simply as $\mathbb{E}[\cdot]$ in the sequel. By expansion, we have
\begin{equation}
\label{Edad}
\mathbb{E}[d_{n,c}^2(d_{n,c}^2+\alpha)]
=\mathbb{E}[d_{n,c}^4]+\alpha \mathbb{E}[d_{n,c}^2].
\end{equation}

Due to the independence of $x_c$, $x_n$, and $y_n$, the second-order moment can be decomposed as
\begin{equation}
\mathbb{E}[d_{n,c}^2]
=\mathbb{E}[(x_n-x_c)^2]+\mathbb{E}[y_n^2]+h^2.
\end{equation}

Since $x_c$ and $x_n$ are i.i.d. uniform random variables over $[0,D_1]$, the variance of their difference can be computed as
\begin{equation}
\mathbb{E}[(x_n-x_c)^2]=\mathrm{Var}(x_n-x_c)=2\mathrm{Var}(x_n)=\frac{D_1^2}{6},
\end{equation}
where we use the fact that $\mathrm{Var}(x_n)=D_1^2/12$ for $x_n\sim U[0,D_1]$. Similarly, for $y_n\sim U[0,D_2]$, the second moment is given by
\begin{equation}
\mathbb{E}[y_n^2]=\mathrm{Var}(y_n)+(\mathbb{E}[y_n])^2=\frac{D_2^2}{12}+\frac{D_2^2}{4}=\frac{D_2^2}{3}.
\end{equation}
Substituting these results yields
\begin{equation}
\mathbb{E}[d_{n,c}^2]=\frac{D_1^2}{6}+\frac{D_2^2}{3}+h^2.
\end{equation}

For the fourth-order moment, we expand $d_{n,c}^4=((x_n-x_c)^2+y_n^2+h^2)^2$ and take expectations term-by-term. Exploiting the independence among $x_n$, $x_c$, and $y_n$, and using the moments $\mathbb{E}[(x_n-x_c)^4]=D_1^4/15$, $\mathbb{E}[y_n^4]=D_2^4/5$, $\mathbb{E}[(x_n-x_c)^2]=D_1^2/6$, and $\mathbb{E}[y_n^2]=D_2^2/3$, we obtain
\begin{equation}
\begin{split}
\mathbb{E}[d_{n,c}^4] =&\ \mathbb{E}[(x_n-x_c)^4]+2\mathbb{E}[(x_n-x_c)^2]\mathbb{E}[y_n^2]\\
&+2h^2\mathbb{E}[(x_n-x_c)^2]+\mathbb{E}[y_n^4]+2h^2\mathbb{E}[y_n^2]+h^4 \\
=&\ \frac{D_1^4}{15}+\frac{D_2^4}{5}+\frac{D_1^2D_2^2}{9}+\frac{h^2D_1^2}{3}+\frac{2h^2D_2^2}{3}+h^4.
\end{split}
\end{equation}
Finally, substituting the expressions for $\mathbb{E}[d_{n,c}^2]$ and $\mathbb{E}[d_{n,c}^4]$ into (\ref{Edad}) completes the derivation of $\mathbb{E}_{x_n,x_c,y_n}[d_{n,c}^2(d_{n,c}^2+\alpha)]$.

Next, we evaluate $\mathbb{E}_{x_c}[d_{n,c}^2(d_{n,c}^2+\alpha)]$ by conditioning on $x_n$ and $y_n$. Let $\delta_n = x_n - D_1/2$ and $\beta_n = y_n^2 + h^2$. Then, the squared distance can be written as
\begin{equation}
d_{n,c}^2 = (\delta_n - \tilde{x}_c)^2 + \beta_n,
\end{equation}
where $\tilde{x}_c = x_c - D_1/2 \sim U[-D_1/2, D_1/2]$. Expanding the expectation and using $\mathbb{E}[\tilde{x}_c^2]=D_1^2/12$ and $\mathbb{E}[\tilde{x}_c^4]=D_1^4/80$, we obtain
\begin{equation}
\begin{split}
\mathbb{E}_{x_c}\!\left[d_{n,c}^2(d_{n,c}^2+\alpha)\right]
=\, & \delta_n^4 + \left(\frac{D_1^2}{2}+2\beta_n+\alpha\right)\delta_n^2 + \frac{D_1^4}{80} \\
& + \frac{(2\beta_n+\alpha)D_1^2}{12} + \beta_n(\beta_n+\alpha).
\end{split}
\end{equation}
This completes the proof.

\bibliographystyle{IEEEtran}
\bibliography{ref}


\vspace{11pt}

\vfill

\end{document}